# Simple visit behavior unifies complex Zika outbreaks

P.D. Manrique and N.F. Johnson

*Department of Physics, University of Miami, Coral Gables, FL 33126, USA*

In addition to sparking the first ever U.S. government health warning for travel within the continental United States, the 2016 Zika virus (ZIKV) outbreak was notable for the geographical resolution with which the three Ground Zero areas were specified, i.e. Wynwood (Fig. 1a), Miami Beach and Little River which are all separated from each other by a short drive within the Miami area. Grubaugh et al. [1] provide a remarkable analysis of ZIKV genomes from infected patients and *Aedes aegypti* mosquitoes, and conclude that multiple introductions of the virus contributed to this 2016 outbreak in Florida; that local transmission likely started several months before its initial detection; and that ZIKV moved among transmission zones in Miami.

However, Ref. 1 provides no proof that these mechanisms can reproduce the human ZIKV outbreak profiles reported in Fig. 1d of Ref. 1 (reproduced in Fig. 1B here) – even though these profiles of human cases represent the key measurement from a scientific and public policy perspective. Our calculations strongly suggest that this is because Ref. 1 fails to consider a simple fact of everyday life: that humans that happen to be in the Miami area may *visit and revisit* these areas on a weekly, if not daily, basis. In particular, these three areas' own residents may by necessity leave and return as part of their daily lives; commuters from outside these areas may enter and then leave; visitors who are resident elsewhere in Miami may visit frequently for social reasons, particularly on weekends; and tourists from outside Miami, whether they be international or not, may visit several times during their stay in Miami.

Figure 1 shows the result of incorporating this simple visit-revisit feature into a minimalistic, generative model (Fig. 1A) that applies equally well to any Ground Zero patch; that scales up to any population and patch size; and that uses only a minimal number of parameters, each of which has a realistic value. The *same* few parameter values explain the complex variations in all three outbreaks in Fig. 1B. Even though the simulations will vary somewhat from run to run, the statistical likelihood of getting such non-monotonic profiles as Fig. 1B from a standard epidemiological model simulation, is negligible ($p \ll 0.05$). The sequential nature of the three profiles in Fig. 1B in calendar time also strongly suggests that there is a simple shift in these visits and revisits from Wynwood over to Miami Beach, then to patch Little River, as Wynwood then Miami Beach became successively declared publicly Ground Zeros. We also note that since *Aedes aegypti* mosquitos are the intermediate carrier between infected and susceptible individuals, our model suggests that the mosquito abundance per trap should follow these temporal profiles nearly exactly, which it does (See Fig. 1d of Ref. 1).

Our visit-revisit model [2, 3] (Fig. 1A) yields an average human flow through the Ground Zero (i.e. average number of individuals entering and leaving) given by $\gamma_{\text{flow}} N$ where $\gamma_{\text{flow}} = 2p_j p_j / (p_j + p_j)$ and $N$ is the total number of humans within the Miami area who wish to visit that Ground Zero. Hence the turnover of individuals within Ground Zero is dictated by $\gamma_{\text{flow}}$ for any $N$. At any timestep, there are on average $\gamma_{\text{occup}} N$ individuals inside Ground Zero who are potentially exposed to infection, where $\gamma_{\text{occup}} = p_j / (p_j + p_j)$ for any $N$. The infection rate for a susceptible individual ($S$) within Ground Zero is $q_i$; the individual recovery rate for an infected individual ($I$) is $q_r$. Here for convenience we take one timestep as one effective day where there is flow of individuals in and out of Ground Zero, though this can easily be generalized. The infection parameters are estimated from official sources recently released [1, 4]. For example, the probability of an individual in Ground Zero getting infected emerges from the percentage of infected mosquitoes collected in pools from June to November 2016 [1]. These findings indicate that one in every 1600 *Aedes aegypti* mosquitoes carried the virus. Hence, the probability of getting infected is $q_i = 0.00014$ per timestep assuming that on average one out of every four people was bitten by a mosquito in a given timestep. The symptoms associated with ZIKV disappear between two and seven days [5] which yields an average recovery probability of $q_r = 0.23$ per timestep. In Fig 1B, $N = 10,000$ of which one quarter occupy the Ground Zero region at a given timestep (i.e. $\gamma_{\text{occup}} = 0.25$) with a flow of 500 new or returning individuals (i.e. $\gamma_{\text{flow}} = 0.05$).

The good agreement for all three Ground Zeros in Fig. 1B means that the parsimonious visit-revisit mechanism in Fig. 1A can be used to inform optimal day-to-day human flow management in order to mitigate an outbreak, in a way that the analysis of Ref. 1 cannot. Figure 2A presents the model's prediction of how changing the human flow will impact the outbreak profile, where the unperturbed scenario is the black curve. This impact is quantified in terms of the outbreak duration and analyzes the results by averaging over 5000 simulations. It predicts that by reducing the flow by 10% during the early stage of the outbreak ($t = 12$ days) the outbreak's duration drops 42% compared to the unperturbed case. Reductions of 30% and 50% during the same early stage reduce the outbreak duration by 79% and 85%, respectively. Later interventions at subsequent stages of the outbreak (pre-peak and post-peak) still help in reducing the outbreak duration, however in smaller percentages (see Fig. 2A right panel). These results point to the need for prompt action to control the human flow and hence maximize the potential benefits.

The visit-revisit mechanism also yields several highly counterintuitive policy implications for future strain variants and for other viruses that cannot be inferred from the analysis in Ref. 1. For example, Fig. 2B illustrates the outbreak severity for more general viruses and parasite-borne diseases [6], according to its recovery rate $q_r$. As can be seen going from left to right panels in Fig. 2B, there is a remarkable nonlinear switch in the dependency of the outbreak severity on increasing human flow through any Ground Zero. For a fast recovery rate (left panel) the severity decreases with increasing flow because the chance of individuals still being infectious when they revisit is now small. By contrast for a slow recovery rate (right panel) the severity increases with increasing flow because many individuals will still be infectious when they revisit. In between, there is a critical value (middle panel) for which the severity increases in a non-monotonic way with increasing flow.

Since 'popular patches' as in Fig. 1A exist on all geographic scales and yet the total flow and occupancy in our visit-revisit model scale directly with $N$, the *same* findings apply in principle whether Ground Zero is a square mile like Wynwood, or an entire city (e.g. Miami) or county (e.g. Miami-Dade), or a state (e.g. Florida), or a country (e.g. U.S.A.) or a region (e.g. the Caribbean) or part of a continent (e.g. South America). In each case, the corresponding results in Fig. 2 hold. In this way, the international country-level travel data presented in Ref. 1 could ultimately be incorporated into a multi-level version of Fig. 1A in which there are patches within patches. The respective public health agencies at each level can therefore use the insight from Fig. 2B to account for the counterintuitive way in which visit-revisit human flow affects the outbreak's duration, severity and time-to-peak at each level of geographical resolution.

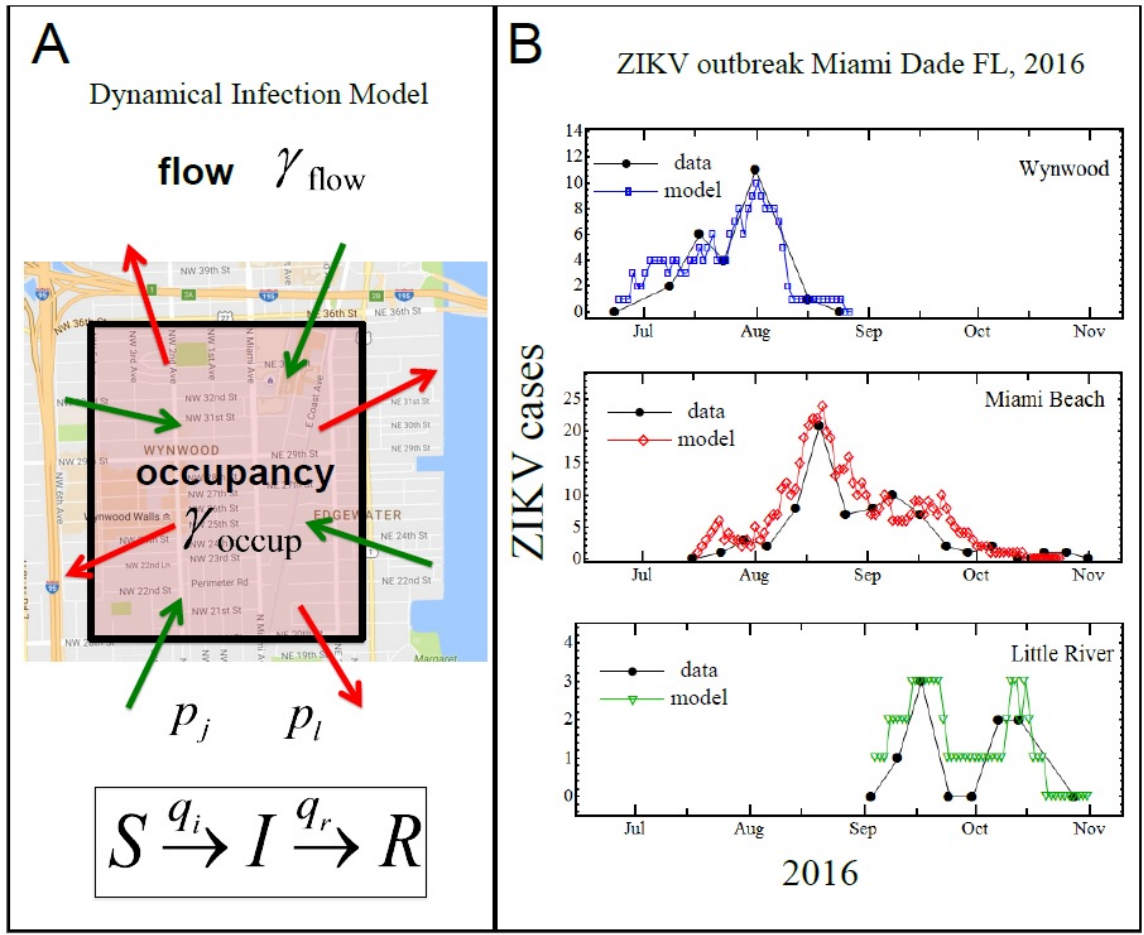

**Figure 1. Unifying visit-revisit mechanism. A**: Ground Zero (red shaded area) and our visit-revisit model parameters. The patch shown corresponds to Wynwood, however the same model applies to all Ground Zero patches, and indeed patches at any geographic scale. Potential visitors to Ground Zero at timestep $t$ enter with probability $p_j$ and leave with probability $p_l$. Their total number is $N$ and susceptible ($S$) individuals within Ground Zero have a probability $q_i$ of getting bitten and becoming infected ($I$) while infected individuals have a probability $q_r$ of recovering ($R$). **B**: Temporal distribution of reported ZIKV cases on a weekly basis for three affected areas in Miami Dade County, FL (Wynwood, Miami Beach, and Little River) during 2016 [1,2] compared to the output of our visit-revisit model. In all three cases, the simulation uses the *same* parameter values, which are all obtained from realistic independent estimates (see text): $q_i = 0.00014$, $q_r = 0.23$, $N = 10,000$, $\gamma_{flow} = 0.05$, and $\gamma_{occup} = 0.25$.

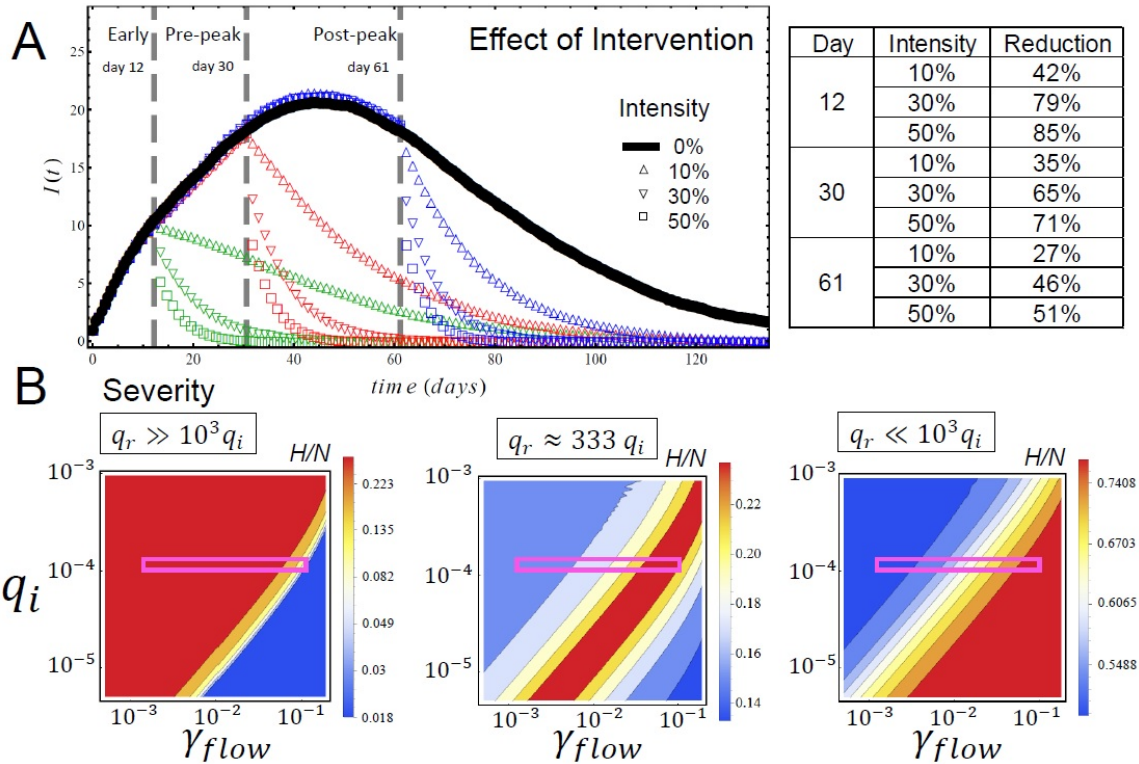

**Figure 2: Predictions and policy implications. A.** Impact of intervention on the infection profile (number of infected individuals). The interventions are applied on different days (vertical gray lines) from the beginning of the outbreak, and with different intensities (symbols). Colors distinguish the times of the intervention: day 12 (green), day 30 (red) and day 61 (blue). The intervention comprises the removal of a specific percentage (10%, 30% and 50%) of *N*, the number of likely visitors. The black thick line is the infection profile without intervention. Each curve is the average profile over 1000 model runs. The end of the outbreak (i.e. duration *T*) is determined by the day on which the average number of infected individuals becomes smaller than one, i.e. $I(t) < 1$. Right panel shows the effect of the intervention on the duration of the outbreak. For each start day of intervention and its intensity, the reduction in the duration is shown with respect to the original unperturbed outbreak. The model parameters are $q_i = 0.00014$, $q_r = 0.23$, $N = 10,000$, $\gamma_{flow} = 0.05$, and $\gamma_{occup} = 0.25$. **B.** Severity (*H/N*) of the outbreak in the scenario that individuals remain infectious for a far shorter time than in **A** (left panel), a far longer time (right panel), and a critical time that divides the two behaviors (middle panel). Pink box points to the contagion probability region associated to the ZIKV outbreak in 2016. In all cases, average individual recovery time $\propto 1/q_r$.

**Author Contribution** P.D.M. and N.F.J. discussed the manuscript. P.D.M. investigated the datasets and performed the model simulations. P.D.M and N.F.J analyzed the results. P.D.M and N.F.J wrote and approved the manuscript.

**Competing Financial Interests** Declared none.